\documentclass[nocopyrightspace,preprint]{sigplanconf}

\usepackage{times}
\usepackage{hyperref}

\usepackage{amsmath,amssymb, latexsym}

\usepackage{amsmath}
\usepackage{amsthm}

\newtheorem{lemma}{Lemma}
\newtheorem{theorem}{Theorem}
\newtheorem*{theorem*}{Theorem}

\newtheorem*{lemma*}{Lemma}

\usepackage{amsfonts}
\usepackage{amssymb}

\usepackage{commands}
\usepackage{liquidHaskell}
\usepackage[inference]{semantic}

\usepackage{enumerate}
\def\url{}
\usepackage{xspace}
\usepackage{epsfig}
\usepackage{booktabs}
\usepackage{listings}
\usepackage{comment}
\usepackage{ifthen}

\newcommand{\isTechReport}{false} 
\newcommand\includeProof[1]{%
  \ifthenelse{\equal{\isTechReport}{true}}
    {{#1}}
    {\ignorespaces}
\xspace}
\newcommand\includeBytestring[1]{}

\def\@envspa{\hspace{0.3em}}
\def\@sa{\hspace{-0.2em}}
\def\@sb{\hspace{0.5em}}
\def\@sc{\hspace{-0.1em}}

{\itshape}{}

\newtheorem*{hypothesis}{Hypothesis}

\newcommand{\proofsketch}{\noindent {\bf Proof Sketch: }}

\usepackage{listings}

\ifdefined\withcolor
	\definecolor{haskellblue}{rgb}{0.0, 0.0, 1.0}
	\definecolor{haskellblue}{rgb}{1.0, 0.0, 0.0}
	\definecolor{gray_ulisses}{gray}{0.55}
	\definecolor{castanho_ulisses}{rgb}{0.71,0.33,0.14}
	\definecolor{preto_ulisses}{rgb}{0.41,0.20,0.04}
	\definecolor{green_ulisses}{rgb}{0.0,0.4,0.0}
\else
	\definecolor{haskellblue}{gray}{0.1}
	\definecolor{haskellred}{gray}{0.1}
	\definecolor{gray_ulisses}{gray}{0.1}
	\definecolor{castanho_ulisses}{gray}{0.1}
	\definecolor{preto_ulisses}{gray}{0.1}
	\definecolor{green_ulisses}{gray}{0.1}
\fi

\def\codesize{\normalsize}

\lstdefinelanguage{HaskellUlisses} {
	basicstyle=\ttfamily\codesize,
	sensitive=true,
	morecomment=[l][\color{gray_ulisses}\ttfamily\codesize]{--},
	morestring=[b]",
	stringstyle=\color{haskellred},
	showstringspaces=false,
	numberstyle=\codesize,
	numberblanklines=true,
	showspaces=false,
	breaklines=true,
	showtabs=false,
	emph=
	{[1]
		FilePath,IOError,abs,acos,acosh,all,and,any,appendFile,approxRational,asTypeOf,asin,
		asinh,atan,atan2,atanh,basicIORun,break,catch,ceiling,chr,compare,concat,concatMap,
		const,cos,cosh,curry,cycle,decodeFloat,denominator,digitToInt,div,divMod,drop,
		dropWhile,either,elem,encodeFloat,enumFrom,enumFromThen,enumFromThenTo,enumFromTo,
		error,even,exp,exponent,fail,filter,flip,floatDigits,floatRadix,floatRange,floor,
		fmap,foldl,foldl1,foldr,foldr1,fromDouble,fromEnum,fromInt,fromInteger,
		fromRational,fst,gcd,getChar,getContents,getLine,head,id,inRange,index,init,intToDigit,
		interact,ioError,isAlpha,isAlphaNum,isAscii,isControl,isDenormalized,isDigit,isHexDigit,
		isIEEE,isInfinite,isLower,isNaN,isNegativeZero,isOctDigit,isPrint,isSpace,isUpper,iterate,
		last,lcm,length,lex,lexDigits,lexLitChar,lines,log,logBase,lookup,map,mapM,mapM_,max,
		maxBound,maximum,maybe,min,minBound,minimum,mod,negate,not,notElem,numerator,odd,
		or,pi,pred,primExitWith,print,product,properFraction,putChar,putStr,putStrLn,quot,
		quotRem,range,rangeSize,read,readDec,readFile,readFloat,readHex,readIO,readInt,readList,readLitChar,
		readLn,readOct,readParen,readSigned,reads,readsPrec,realToFrac,recip,rem,repeat,replicate,
		reverse,round,scaleFloat,scanl,scanl1,scanr,scanr1,seq,sequence,sequence_,show,showChar,showInt,
		showList,showLitChar,showParen,showSigned,showString,shows,showsPrec,significand,signum,sin,
		sinh,snd,span,splitAt,sqrt,subtract,succ,sum,tail,take,takeWhile,tan,tanh,threadToIOResult,toEnum,
		toInt,toInteger,toLower,toRational,toUpper,truncate,uncurry,undefined,unlines,until,unwords,unzip,
		unzip3,userError,words,writeFile,zip,zip3,zipWith,zipWith3,listArray,doParse,for,initTo,
        maxEvens,create,get,set,initialize,idVec,fastFib,fibMemo,
        insert,union,split,size,fromList,initUpto,trim,quickSort,insertSort,append,upperCase,
        copy, group, doDownLoop, mapAccumR, peekByteOff,
        pokeByteOff,spanByte, 
        good, bad, foo, explode, 
        fib, ack, 
        tLen,
        memcpy,writeChar,unsafeWrite,unsafeFreeze,
        singleton
	},
	emphstyle={[1]\color{haskellblue}},
	emph=
	{[2]
		Bool,Char,Double,Either,Float,IO,Integer,Int,Maybe,Ordering,Rational,Ratio,ReadS,ShowS,String,
		Word8,Nat,NonZero,Nat64,Text,ByteString,ByteStringSZ,ByteStringN,
        Ptr,ForeignPtr,CSize
        InPacket,Tree,Prop,TreeEq,TreeLt,Vec,
        NullTerm,IncrList,DecrList,UniqList,BST,MinHeap,MaxHeap,
        PtrN,ByteStringN,ByteStringEq,VO,ByteStringsEq,ByteStringNE
	},
	emphstyle={[2]\color{castanho_ulisses}},
	emph=
	{[3]
		case,class,data,deriving,do,else,if,return,def,import,in,infixl,infixr,instance,let,
		module,measure,predicate,of,primitive,then,refinement,type,where
	},
	emphstyle={[3]\color{preto_ulisses}\textbf},
	emph=
	{[4]
		quot,rem,div,mod,elem,notElem,seq
	},
	emphstyle={[4]\color{castanho_ulisses}\textbf},
	emph=
	{[5]
		PS,Tip,Node,EQ,False,GT,Just,LT,Left,Nothing,Right,True,Show,Eq,Ord,Num
	},
	emphstyle={[5]\color{green_ulisses}}
}

\lstnewenvironment{code}
{\lstset{language=HaskellUlisses}}
{}

\lstMakeShortInline[language=HaskellUlisses]@

\begin{document}

\title{From Safety To Termination And Back: SMT-Based Verification For Lazy Languages}



\authorinfo{Niki Vazou \and Eric L. Seidel \and Ranjit Jhala}
           {UC San Diego}{}

\maketitle


\begin{abstract}
SMT-based verifiers have long been an effective means of 
ensuring safety properties of programs.
While these techniques are well understood, we show that they 
implicitly require \emph{eager} semantics; directly applying 
them to a \emph{lazy} language is unsound due to the presence 
of divergent sub-computations.
We recover soundness by composing the safety analysis with a
\emph{termination} analysis.
Of course, termination is itself a challenging problem, but 
we show how the safety analysis can be used to ensure termination, 
thereby bootstrapping soundness for the entire system.
Thus, while safety invariants have long been required to 
prove termination, we show how termination proofs can be 
to soundly establish safety.
We have implemented our approach in \toolname, a Refinement 
Type-based verifier for Haskell.
We demonstrate its effectiveness via an experimental evaluation 
using \toolname to verify safety, functional correctness and 
termination properties of real-world Haskell libraries, totaling
over 10,000 lines of code.


\end{abstract}

\section{Introduction}\label{sec:intro}

SMT-based verifiers, based on Floyd-Hoare Logic 
(\eg EscJava~\cite{ESCJava}),
or combined with abstract interpretation (\eg SLAM~\cite{BMMR01}),  
have been highly effective at the automated verification of imperative programs.
In the functional setting, these techniques are generalized as  
\emph{refinement types}, where invariants are encoded
by composing types with SMT-decidable refinement 
predicates~\cite{Rushby98,pfenningxi98}.
For example
\begin{code}
   type Pos = {v:Int | v >  0}
   type Nat = {v:Int | v >= 0}
\end{code}
are the basic type @Int@ refined with logical predicates that state that ``the values" @v@ described
by the type are respectively strictly positive and non-negative.
We encode \emph{pre-} and \emph{post-}conditions (contracts) using 
refined function types like 
\begin{code}
   div :: n:Nat -> d:Pos -> {v:Nat | v <= n}
\end{code}
which states that the function @div@ \emph{requires} inputs that are 
respectively non-negative and positive, and \emph{ensures} that the 
output is less than the first input @n@.
If a program containing @div@ statically type-checks, we can rest assured that
executing the program will not lead to any unpleasant divide-by-zero errors.
%
%
Several groups have demonstrated that refinements can be used to statically 
verify properties ranging from simple array safety~\cite{pfenningxi98,LiquidPLDI08}
to functional correctness of data structures~\cite{LiquidPLDI09}, 
security protocols~\cite{GordonTOPLAS2011,Bhargavan13},  
and compiler correctness~\cite{SwamyPOPL13}.

Given the remarkable effectiveness of the technique, we embarked
on the project of developing a refinement type based verifier for 
Haskell, assuming that the standard soundness proofs from 
Floyd-Hoare logics and refinement types would carry over directly.
Of course, the previous logics and systems were all developed 
for eager, \emph{call-by-value} languages, but 
we presumed 
that the order of evaluation would surely prove irrelevant, and 
that the soundness guarantees would translate to Haskell's
lazy, \emph{call-by-need} regime.

To our surprise, we were totally wrong. 

\spara{1. Laziness Precludes Partial Correctness}
Our first contribution is to demonstrate that refinement typing
is \emph{unsound} in the presence of lazy evaluation~(\S~\ref{sec:overview}).
Consider the program:
%

\begin{code}
foo     :: n:Nat -> {v:Nat | v < n}
foo n   = if (n > 0) then (n-1) else (foo n)

bar     :: z:Pos -> x:Int -> Int
bar z x = 2014 `div` z

main    = let (a,b) = (0, foo 0) in bar a b
\end{code}

A standard refinement type checker will happily verify the above program. 
The refinement type signature for @foo@ captures the \emph{partial correctness}
property: the function @foo@ requires non-negative inputs and ensures that its 
output (\emph{if} one is produced!) will be strictly less than its input.
Consequently, the checker concludes that at the call-site for @bar@, the value @z@
equals $0$ and the value @x@ is some non-negative integer that is strictly less than $0$.
In other words, the checker concludes that the environment is \emph{inconsistent} 
and hence trivially type checks.

The issue is independent of refinement typing and affects any Floyd-Hoare
logic-based verifier. For example, a hypothetical ESC-Scala would verify the
following code which restates the above using classical 
@requires@ and @ensures@ clauses: 
\begin{code}
def foo(n:Int):Int =
  //@ requires (0 <= n)
  //@ ensures  (0 <= \result && \result < n)
  if (n > 0) return (n-1) else return foo(n) 
  
def bar(z:Int, x:=>Int) = 
  //@ requires 0 < z
  return (2014 div z)
  
def main = {val (a,b) = (0,foo(0));bar(a,b)} 
\end{code}

One should not be alarmed as this deduction is perfectly sound under 
eager, \emph{call-by-value} semantics. In both cases, the verifier
determines that the call to @bar@ is \emph{dead code} -- the call 
is safe because it is not invoked at all.
This reasoning is quite unsound for Haskell's lazy, \emph{call-by-need}
semantics, and Scala's lazy, \emph{call-by-name} parameters (indicated
by the @:=>@ type annotation). In both cases, 
the program execution
would skip blithely over the call to @foo@, plunge headlong into the
@div@, and crash.

As we show, the problem is that with lazy evaluation, one can only 
trust a refinement or invariant if one is guaranteed that evaluating 
the corresponding term will \emph{not diverge}.
That is, the classical Floyd-Hoare separation of ``partial" and ``total"
correctness breaks down, and even safety verification requires checking termination.

\spara{2. Termination for and by Refinement Typing}
The prognosis seems dire: to solve one problem it appears we must first solve a harder one!
Our second contribution is to demonstrate that refinement types can themselves be 
used to prove termination~(\S~\ref{sec:overview}).
In particular, we show how to adapt the classical idea of ranking functions~\cite{Turing36}, 
as embodied via sized types~\cite{HughesParetoSabry96,BartheTermination}, to the setting of
Refinement Typing. 
The key idea is to ensure that each recursive call is made with parameters of strictly 
decreasing @Nat@-valued size.

We show that refinements naturally encode sized types and generalize them in useful
ways by allowing
(1)~different notions of size to account for recursive data types,
(2)~lexicographically ordered ranking functions to support complex forms of recursion, and
    most importantly
(3)~the use of auxiliary relational invariants (circularly, via refinement types!) 
    to verify that sizes decrease in non-structurally recursive functions.

Our use of refinements to prove termination makes proving soundness interesting in two ways. 
First, to check Haskell code-bases, we cannot require that \emph{every} term terminates, 
and so we must support programs containing terms that may diverge.
Second, there is a circularity in the soundness proof itself as termination is 
required to prove refinements sound and vice versa.
We address these issues by developing a core calculus~(\S~\ref{sec:language}) with
\emph{optimistic} semantics~\cite{Ennals03}, proving those semantics equivalent to 
call-by-name evaluation, and then proving soundness with respect to the optimistic 
semantics~(\S~\ref{sec:typing}). 
Thus, while it is well known that safety properties (invariants) are
needed to prove termination~\cite{Terminator}, we show for the
first time how termination properties are needed to prove safety.

\spara{3. Refinement Types for Real-World Haskell}
We have implemented our technique by extending \toolname~\cite{vazou13}.
Our third contribution is an experimental evaluation of the effectiveness 
of our approach by using \toolname to check substantial, real-world Haskell 
libraries 
totaling over $10,000$ lines of code~(\S~\ref{sec:evaluation}).
The verification of these libraries requires precisely analyzing recursive
structures like lists and trees, tracking the relationships between their
contents, inferring invariants in the presence of low-level pointer arithmetic,
and lifting the analysis across polymorphic, higher-order functions.
We demonstrate that by using \toolname we were able to prove termination and a
variety of critical safety and functional correctness properties with a modest
number of manually specified hints, and even find and fix a subtle correctness 
bug related to unicode handling in \libtext.
To the best of our knowledge, this is the most substantial evaluation of
refinement types on third party code, and demonstrates that SMT-based safety-
and termination-verification can be practical and effective for higher-order, 
functional languages.



\section{Overview}\label{sec:overview}

We start with an overview of our contributions.
After quickly recapitulating the basics of 
\emph{refinement types} we illustrate why
the approach is unsound in the presence of 
lazy evaluation.  
Next, we show that we can recover soundness by coupling 
refinements with a termination analysis. 
Fortunately, we demonstrate how we can bootstrap off of
refinements to solve the termination problem, and 
hence obtain a sound and practical verifier for Haskell.

\spara{SMT-Based Refinement Type Checking} 
Recall the refinement type aliases @Pos@ and @Nat@ 
and the specification for @div@ from \S~\ref{sec:intro}.
A refinement type system will use these to \emph{reject} 
\begin{code}
    bad :: Nat -> Nat -> Int
    bad x y = x `div` y
\end{code}
because, to check that the second parameter @y@ has type @Pos@
at the call to @div@, the system will issue a subtyping query
\begin{align*}
\ttbind{x}{\ttref{x \geq 0}},\ \ttbind{y}{\ttref{y \geq 0}}\ \vdash\ & \subtref{v=y}{v
> 0}
\intertext{which reduces to the \emph{invalid} SMT query}
   \mathtt{\mathtt{(x \geq 0)}} \wedge \mathtt{(y \geq 0)}\ \Rightarrow\ & \mathtt{(v = y)}
   \Rightarrow \mathtt{(v > 0)}
\end{align*}
On the other hand, the system will \emph{accept} the program
\begin{code}
    good :: Nat -> Nat -> Int
    good x y = x `div` (y + 1)
\end{code}
Here, the corresponding subtyping query is
\begin{align}
\ttbind{x}{\ttref{x \geq 0}},\  
\ttbind{y}{\ttref{y \geq 0}}\ \vdash\ & \subtref{v=y+1}{v > 0}
\label{sub:good}
\intertext{which reduces to the \emph{valid} SMT query}
   \mathtt{(x \geq 0)} \wedge \mathtt{(y \geq 0)}\ \Rightarrow\ & \mathtt{(v = y+1)}
   \Rightarrow \mathtt{(v > 0)}
\end{align}

\subsection{Laziness Makes Refinement Typing Unsound}

Next, let us look in detail at the program with @foo@ and @bar@
from \S~\ref{sec:intro}.
A standard refinement type checker first verifies 
the signature of @foo@ in the classical rely-guarantee 
fashion, by (inductively) assuming its type and checking
that its body yields the specified output. 
In the @then@ branch, the output subtyping obligation
\begin{align*}
\ttbind{n}{\ttref{n \geq 0}},\ \ttbind{\_}{\ttref{n > 0}} \ \vdash\ & 
   \subtref{v = n-1}{0 \leq v \land v < n}
\intertext{reduces to the valid SMT formula} 
\mathtt{n \geq 0}\ \wedge\ \mathtt{n > 0} \ \Rightarrow\ & 
   \mathtt{(v = n-1)} \Rightarrow\ \mathtt{(0 \leq v \land v < n)}
\end{align*}
In the @else@ branch, the output is proven by inductively 
using the assumed type for @foo@.
Next, inside @main@ the binder @b@ is assigned the 
output type of @foo@ with the formal @n@ replaced with 
the actual @0@. 
Thus, the subtyping obligation at the call to @bar@ is
\begin{align*}
\ttbind{a}{\ttref{a = 0}},\ \ttbind{b}{\ttref{0 \leq b \wedge b < 0}} \ \vdash\ & 
   \subtref{v = a}{v > 0}
\intertext{which reduces to the SMT query}
\mathtt{a = 0}\ \wedge\ \mathtt{(0 \leq b \wedge b < 0)}\ \Rightarrow\ &
   \mathtt{(v = a)} \Rightarrow \mathtt{(v > 0)}
\end{align*}
which is trivially valid as the antecedent is inconsistent.

Unfortunately, this inconsistency is unsound under Haskell's lazy evaluation! 
Since @b@ is not required, the program will dive headlong into evaluating 
the @div@ and hence crash, rendering the type checker's guarantee meaningless.

\spara{Reconciling Laziness And Refinements}
One may be tempted to get around the unsoundness 
via several different routes.
First, one may be tempted to point the finger of 
blame at the ``inconsistency" itself. 
Unfortunately, this would be misguided, since 
such inconsistencies are not a bug but a crucial 
feature of refinement type systems. 
They enable, among other things, \emph{path sensitivity}
by incorporating information from run-time tests (guards)
and hence let us verify that expressions that throw
catastrophic exceptions (\eg @error e@)
are indeed unreachable dead code and will not 
explode at run-time.
Second, one might use a CPS transformation \cite{PlotkinTCS75,WadlerICFP03} to 
convert the program into call-by-value.
We confess to be somewhat wary of the prospect of translating inferred types and errors 
\emph{back} to the source level after such a transformation.
Previous experience shows that the ability to map types and errors to source is critical 
for usability.
Third, one may want some form of \emph{strictness analysis}~\cite{Mycroft80} 
to statically predict which expressions must be evaluated, and only use refinements
for those expressions. 
This route is problematic as it is unclear whether one can develop a sufficiently precise
strictness analysis.
More importantly, it is often useful to add \emph{ghost} values into the program for the 
sole purpose of making refinement types \emph{complete}~\cite{TerauchiPOPL13}. 
By construction these values are not used by the program, and would be thrown away
by a strictness analysis, thus precluding verification.

\subsection{Ensuring Soundness With Termination}

The crux of the problem is that when we establish that 
\begin{align*}
        & \hastype{\Env}{\tte}{\ttreft{\ttv}{\ttInt}{\ttp}} 
\intertext{what we have guaranteed is that
\emph{if} $\tte$ reduces to an integer $n$, 
\emph{then} $n$ satisfies the 
logical predicate $\ttp\sub{\ttv}{n}$~\cite{Knowles10,Greenberg12}.
Thus, to account for diverging computations, we should properly 
view the above typing judgment as weakened with a \emph{bottom disjunct}
}
    & \hastype{\Env}{\tte}{\ttreft{\ttv}{\ttInt}{v = \perp \vee\ \ttp}}
\end{align*}

Now, consider the expression @let x = e in e'@.
In an eager setting, we can readily eliminate the bottom disjunct and assume that 
@x@ satisfies @p[x/v]@ when analyzing @e'@ because if @e@ diverges, then @e'@ 
is \emph{not evaluated}.
In other words, the mere fact that evaluation of @e'@ \emph{began} allows us 
to conclude that ${\ttx \not = \perp}$, and so we can eliminate the bottom disjunct
without compromising soundness.
However, in a lazy setting we cannot drop the bottom disjunct because we may well 
evaluate @e'@ even if @e@ diverges!

One way forward is to take the bull by the horns and \emph{directly} reason
about divergence and laziness using the bottom disjunct. That is, to \emph{weaken}
each refinement 
with the bottom disjunct. 
While sound, such a scheme is imprecise as the hypotheses will be too weak to let us
prove interesting \emph{relational} invariants connecting \emph{different} program 
variables. For instance, the subtyping query for @good@ (\ref{sub:good}),
if we ignore the @x@ binder, becomes:
%
\begin{align*}
\ttbind{y}{\ttref{y = \perp \vee\ y \geq 0}}\ \vdash\ & 
	\subtref{v = \perp\vee\ v=y+1}{v = \perp\vee\ v > 0}
\intertext{which boils down to the SMT query}
   \mathtt{(y = \perp \vee\ y \geq 0)}\ \Rightarrow\ & \mathtt{(v = \perp\vee\ v = y+1)}
   \Rightarrow \mathtt{(v = \perp\vee\ v > 0)}
\end{align*}
which is invalid, causing us to reject a perfectly safe program!

One might try to make the direct approach a little less na\"{i}ve by somehow
axiomatizing the semantics of operators like @+@ to stipulate that the result
(\eg @v@ above) is only non-bottom when the operands (\eg @y@ and @1@) are
non-bottom. 
In essence, such an axiomatization would end up encoding 
lazy evaluation inside the SMT solver's logic, and would indeed make
the above query valid. However, it is quite unclear to us how to design
such an axiomatization in a systematic fashion. Worse, even with such an
axiomatization, we would never be able to verify trival programs like
\begin{code}
baz :: x:Int -> y:Int -> {z:Int | x > y}
    -> Int
baz x y z = assert (x > y) 0
\end{code}
as the bottom disjunct on @z@ -- which is never evaluated -- would 
preclude the verifier from using the refinement relating the values 
of @x@ and @y@ that is needed to prove the assertion.
The above example is not contrived; it illustrates a common idiom 
of using ghost variables that carry ``proofs" about other program 
variables.

The logical conclusion of the above line of inquiry is that to
restore soundness and preserve precision, we need a means of 
precisely eliminating the bottom disjunct, \ie of determining 
when a term is definitely \emph{not} going to diverge. 
That is, we need a termination analysis.  
With such an analysis, using Reynolds'~\cite{Reynolds72} terminology,
we could additionally type each term as either 
\trivial, meaning it \emph{must} terminate, or
\serious, meaning it \emph{may not} terminate.
Furnished with this information, the refinement type checker may 
soundly drop the bottom disjunct for \trivial expressions, and 
keep the bottom disjunct (or just enforce the refinement @true@)
for \serious terms.


Our approach is properly viewed as an optimization 
that strengthens the ``direct" refinement (with a bottom disjunct) 
with a termination analysis that lets us eliminate the
bottom disjunct in the common case of terminating terms.
In general, this approach leaves open the possibility of directly 
reasoning about bottom (\eg when reasoning about infinite streams),
as we \emph{do not} require that \emph{all} terms be 
provably terminating, but only the ones with (non-trivial) refinements 
\emph{without} the bottom disjunct. 
The dual approach -- using safety invariants to strengthen 
and prove termination is classical~\cite{Terminator}; this 
is the first time termination has been used to strengthen 
and prove safety!

As an aside, readers familiar with fully dependently typed languages 
like Agda~\cite{norell07} and Coq~\cite{coq-book} may be unsurprised 
at the termination requirement.  
However, the role that termination plays here is quite different.
In those settings, arbitrary terms may appear in types; termination ensures the
semantics are well-defined and facilitates type equivalence.
In contrast, refinement logics are carefully designed to preclude arbitrary
terms; they only allow logical predicates over well-defined, decidable theories,
which crucially \emph{also} include program variables.
As we saw, this choice is sound under call-by-value, but problematic under
call-by-name, as in the latter setting even a mere first-order variable can
correspond to an undefined diverging computation.

\subsection{Ensuring Termination With Refinements}\label{sec:overview:termination}

How shall we prove termination?
Fortunately, there is a great deal of research on this problem. 
The heart of almost all the proposed solutions is the classical
notion of \emph{ranking functions}~\cite{Turing36}: in any 
potentially \emph{looping} computation, prove that some 
well-founded metric strictly decreases every time around the loop.

\spara{Sized Types}
In the context of typed functional languages, the primary
source of looping computations is recursive functions. 
Thus, the above principle can be formalized by associating
a notion of \emph{size} with each type, and verifying that 
in each recursive call, the function is invoked with 
arguments whose size is \emph{strictly smaller} than 
the current inputs.
Thus, one route to recovering soundness would be to 
perform a first phase of size analysis 
like~\cite{HughesParetoSabry96,Sereni05,BartheTermination} 
to verify the termination of recursive functions, and then
carry out refinement typing in a second phase. 
However, (as confirmed by our evaluation) proving that sizes
decrease often requires auxiliary invariants of the kind 
refinements are supposed to establish in the first 
place~\cite{Terminator}.
Instead, like~\cite{XiTerminationLICS01}, we develop a means
of encoding sizes and proving termination circularly, via 
refinement types.

\spara{Using The Value As The Size}
Consider the function 
\begin{code}
    fib  :: Nat -> Int
    fib 0 = 1
    fib 1 = 1
    fib n = fib (n-1) + fib (n-2)
\end{code}
Recall that in our setting @Nat@ is simply non-negative
@Int@eger.
Thus, we can naturally associate a size with
its \emph{own value}. 

\spara{Termination via Environment-Weakening}
To verify \emph{safety}, a standard refinement type 
checker (or Hoare-logic based program verifier) would
check the body \emph{assuming} an environment: 
\begin{align*}
\ttbind{\ttn}{\ttNat},\ 
\ttbind{\ttfib}{\ttNat \rightarrow \ttInt} &  
\intertext{Consequently, it would verify that at the 
recursive call-site the argument $\mathtt{n-2}$ is
a $\ttNat$, by checking the SMT validity of}
\mathtt{(n \geq 0 \wedge\ n \not = 0 \wedge n \not = 1)} \Rightarrow\
\mathtt{(v = n-2)} \Rightarrow\ & \mathtt{v \geq 0} \\
\intertext{and the corresponding formula for $\mathtt{n-1}$, 
thereby guaranteeing that $\ttfib$ respects its signature.
For \emph{termination}, we tweak the procedure 
by checking the body in a \emph{termination-weakened environment}}
\ttbind{\ttn}{\ttNat},\ 
\ttbind{\ttfib}{\ttreft{\ttnp}{\ttNat}{\ttnp < \ttn} \rightarrow \ttInt} &
\intertext{where we have weakened the type of $\ttfib$ by stipulating 
that it \emph{only} be recursively called with $\ttNat$ values $\ttnp$ 
that are \emph{strictly less than} the current parameter $\ttn$. 
The body still type checks as}
\mathtt{(n \geq 0 \wedge\ n \not = 0 \wedge n \not = 1)}\ 
\Rightarrow\ \mathtt{(v = n-2)} 
\Rightarrow\ & 
\mathtt{v \geq 0 \wedge v < n}
\end{align*}
is a valid SMT formula. 
We prove (Theorem~\ref{th:termination})
that since the body typechecks under 
the weakened assumption for the 
recursive binder, the function will
terminate on all $\ttNat$s.

\spara{Lexicographic Termination}
Our environment-weakening technique generalizes
to include so-called lexicographically decreasing measures.
For example, consider the Ackermann function.
\begin{code}
ack :: Nat -> Nat -> Nat
ack m n 
  | m == 0    = n + 1
  | n == 0    = ack (m-1) 1 
  | otherwise = ack (m-1) (ack m (n-1))
\end{code}
We cannot prove that some argument always decreases;
the function terminates because \emph{either} the 
first argument strictly decreases \emph{or} the 
first argument remains the same and the second argument 
strictly decreases, \ie the pair of arguments strictly
decreases according to a well-founded lexicographic 
ordering.
To account for such functions, we generalize the 
notion of weakening to allow a sequence of witness arguments 
while requiring that 
(1)~each \emph{witness} argument be \emph{non-increasing}, and
(2)~the \emph{last witness} argument be \emph{strictly decreasing} 
if the previous arguments were equal.
These requirements can be encoded by generalizing the 
notion of environment-weakening: we check the body of
@ack@ under
\begin{align*}
\ttm & : \ttNat,\\
\ttn & : \ttNat,\\
\ttack & : {\ttmp\colon\ttNat \rightarrow 
            \ttreft{\ttnp}{\ttNat}{\ttmp \leq \ttm \lor \ttmp = \ttm \Rightarrow \ttnp < \ttn} \rightarrow 
            \ttNat}
\end{align*}
Thus, while sound refinement typing requires proving 
termination, on the bright side, refinements make 
proving termination easy.

\spara{Measuring The Size of Structures}
Consider the function @map@ defined over the standard list type.
\begin{code}
map f []        = [] 
map f ys@(x:xs) = f x : map f xs
\end{code}
In @map@, the recursive call is made to a ``smaller" input.
We formalize the notion of size with \emph{measures}.
\begin{code}
measure len :: [a] -> Nat
len []      = 0
len (x:xs') = 1 + (len xs')
\end{code}
With the above definition, 
the @measure@ strengthens the type of the data constructors to:
\begin{code}
[]  :: {v: [a] | len v = 0}
(:) :: x:a -> xs:[a]
    -> {v:[a] | len v = 1 + len xs}
\end{code}
where @len@ is simply an uninterpreted function in SMT logic~\cite{LiquidPLDI09}.
%
We can now verify that @map@ does not change the @len@gth of the list: 
\begin{code}
type ListEq a YS = {v:[a] | len v = len YS}

map :: (a -> b) -> ys:[a] -> ListEq b ys
\end{code}
This type is only valid if @map@ provably terminates.
We simultaneously verify termination and the type as
before, by checking the body in the termination-weakened
environment
\renewcommand\ttt{\mathtt{ys}}
\renewcommand\tttp{\mathtt{ys'}}
\renewcommand\ttsz{\mathtt{len}}
\begin{align*}
\mathtt{ys} & : \mathtt{[a]}\\
\mathtt{map} & : \mathtt{(\tta \rightarrow \ttb)} \rightarrow 
           \ttreft{\tttp}{\mathtt{[a]}}{{len}\ \tttp < {len}\ \ttt} \rightarrow
           \mathtt{ListEq}\ \ttb\ \tttp
\end{align*}
To ensure that the body recursively uses @map@ per its
weakened specification, 
the recursive call @map f xs@ generates the subtyping query
\begin{align*}
\mathtt{ys}  & : \ttref{{len}\ ys = 1 + {len}\ xs}  \\
\mathtt{xs} & : \ttref{{len}\ xs \geq 0} \ \vdash\ \subt{\ttref{\tttp = xs}}{\ttref{\ttsz\ \tttp < \ttsz\ ys }}\\
\intertext{Thanks to the $\mathtt{case}$ unfolding, the type of $\mathtt{ys}$ is strengthened with the measure 
relationship with the tail $\mathtt{xs}$~\cite{LiquidPLDI09}.
Hence, the subtyping above reduces to the valid SMT query}
        & \mathtt{{len}\ ys = 1 + {len}\ xs}  \\
\wedge\ & \mathtt{{len}\ xs \geq 0} \ \Rightarrow\ \mathtt{(\tttp = xs)}\
\Rightarrow\ \mathtt{({len}\ \tttp < {len}\ ys)}
\end{align*}
Hence, using refinements and environment-weakening, we simultaneously 
verify termination and the output type for @map@.

\spara{Witnessing Termination}
Sometimes, the decreasing metric cannot be associated with a single parameter or 
lexicographically ordered sequence of parameters, but is instead an auxiliary 
value that is a \emph{function} of the parameters.
For example, here is the standard @merge@ function from the eponymous sorting
procedure:
\begin{code}
merge xs@(x:xs') ys@(y:ys') 
  | x < y     = x : merge xs' ys
  | otherwise = y : merge xs  ys'
\end{code}
neither parameter provably decreases in both calls, but the \emph{sum}
of the sizes of the parameters strictly decreases. 

In these cases, we just explicate the metric with a \emph{ghost} parameter
that acts as a witness for termination:
\begin{code}
merge d xs@(x:xs') ys@(y:ys') 
  | x < y     = x : merge dx xs' ys
  | otherwise = y : merge dy xs  ys'
  where 
    dx        = length xs' + length ys
    dy        = length xs  + length ys'
\end{code}
where @length :: zs:[a] -> {v:Nat | v = len zs}@ returns the number of elements in the list.
Now, the system verifies that @d@ equals the sum of the sizes of the two lists, and 
that in each recursive call @dx@ and @dy@ are strictly smaller than @d@, thereby proving 
that @merge@ terminates.

\includeProof{\input{overview.fin}}

\section{Language}\label{sec:language}

Next, we present a core calculus \corelan that
formalizes our approach of refinement types 
under lazy evaluation.
Instead of proving soundness directly 
on lazy, \emph{call-by-name} (CBN)
semantics, we prove soundness with respect
to an \emph{optimistic} (OPT) semantics where
(provably) terminating terms are \emph{eagerly}
evaluated, and we separately prove an equivalence
relating the CBN and OPT evaluation strategies.
With this in mind, let us see the syntax (\S~\ref{sec:syntax}), 
the dynamic semantics (\S~\ref{sec:semantics}), 
and finally, the static semantics (\S~\ref{sec:typing}).

\subsection{Syntax}\label{sec:syntax}
Figure \ref{fig:syntax} summarizes the syntax 
of expressions and types.

\spara{Constants}
The primitive constants include basic values like
$\mathtt{true}$, 
$\mathtt{false}$, 
$\mathtt{0}$, 
$\mathtt{1}$, \etc,
arithmetic and logical operators like
$\mathtt{+}$,$\mathtt{-}$,$\mathtt{\leq}$,$\mathtt{/}$, 
$\mathtt{\land}$, $\mathtt{\lnot}$.
In addition, we include a special (untypable)
$\CRASH$ constant that models errors.
Primitive operations will return a $\CRASH$
when invoked with inputs outside their domain, 
\eg when $\mathtt{/}$ is invoked with a $\mathtt{0}$ 
divisor, or an $\mathtt{assert}$ is called with $\mathtt{false}$.

\spara{Expressions} In addition to the primitive
constants $c$, \corelan expressions include 
the variables $x$, 
$\lambda$-abstractions $\efun{x}{\tau}{e}$
and applications $\eapp{e}{e}$,
let-binders $\elet{x}{e}{e}$, and
the fix operator $\erec{f}{x}{e}$ 
for defining potentially diverging
recursive functions.
%

\begin{figure}[t!]
\centering
$$
\begin{array}{rrcl}
\emphbf{Expressions} \quad 
  & e 
  & ::= 
  &      x 
  \spmid c 
  \spmid \efun{x}{\tau}{e} 
  \spmid \eapp{e}{e}
  \\ && \spmid & 
  		 \elet{x}{e}{e} 
  \spmid \erec{f}{x}{e} 
  \\[0.05in] 

\emphbf{Basic Types} \quad 
  & b 
  & ::= 
  &      \tint
  \spmid \tbool
  \\[0.05in]

\emphbf{Label} \quad 
  & l
  & ::= 
  &      \lbot
  \spmid \ltop
  \\[0.05in]

\emphbf{Type} \quad 
  & \tau
  & ::= 
  &      
     \tref{v}{b}{l}{e} \spmid \tfun{x}{\tau}{\tau}
\end{array}
$$
\caption{Syntax of \corelan Terms and Types}
\label{fig:syntax}
\end{figure}

\subsection{Dynamic Semantics}\label{sec:semantics}

\begin{figure}[t!]
$$
\begin{array}{rrcl}
\emphbf{Contexts} \quad 
  & C
  & ::= 
  &      \bullet 
  \spmid \eapp{v}{C} 
  \spmid \elet{x}{C}{e} 
  \\[0.05in] 

\emphbf{Values} \quad 
  & v
  & ::= 
  & c 
  \spmid \efun{x}{\tau}{e} 
  \spmid \erec{f}{x}{e} 
  \\[0.05in] 
\end{array}
$$


\setlength\arraycolsep{3pt}
$$
\begin{array}{rclll}
\eval{C[e]&}{&C[e']} & \text{if}\ \FORCE{e}, \eval{e}{e'}& \\
\eval{\eapp{(\efun{x}{\tau_x}{e})}{e_x}&}{&\SUBST{e}{x}{e_x}}
& \text{if}\ \NOFORCE{e_x} &\\
\eval{\eapp{(\erec{f}{x}{e})}{e_x}&}{&\SUBST{\SUBST{e}{x}{e_x}}{f}{\erec{f}{x}{e}}} 
& \text{if}\ \NOFORCE{e_x} &\\
\eval{\elet{x}{e_x}{e}&}{&\SUBST{e}{x}{e_x}}                                    
& \text{if}\ \NOFORCE{e_x} &\\
\eval{\eapp{e_1}{e_2}&}{&\eapp{e_1'}{e_2}}
& \text{if}\ \eval{e_1}{e_1'} &\\
\eval{\eapp{c}{v}&}{&\dbrkts{c}(v)}
\end{array}
$$
\caption{Operational Semantics of \corelan}
\label{fig:semantics}
\end{figure}
We define the dynamic behavior of \corelan programs
in Figure \ref{fig:semantics} using a small-step 
contextual operational semantics. 


\spara{Forcing Evaluation}
In our rules, contexts and values are defined as usual.
In addition to the usual lazy semantics, our soundness 
proof requires a ``helper" optimistic semantics.
Thus, we parameterize the small step rules with a 
\emph{force predicate} $\FORCE{e}$ that is used to 
determine whether to force evaluation of the 
expression $e$ or to defer evaluation until 
the value is \emph{needed}.

\spara{Evaluation Rules}
We have structured the small-step rules 
so that if a function parameter or 
let-binder $\FORCE{e}$ then we eagerly 
force evaluation of $e$ (the first rule), 
and otherwise we (lazily) substitute the 
relevant binder with the unevaluated 
parameter expression (the second, third and fourth rule).
The fifth rule evaluates the function in an application, 
as its value is always needed. 

\spara{Constants}
The final rule, application of a
constant, requires the argument be 
reduced to a value; in a single step
the expression is reduced to the 
output of the primitive constant operation.

\spara{Eager and Lazy Evaluation}
To understand the role played by $\R$, 
consider an expression of the form 
$(\efun{x}{\tau_x}{e})\ e_x$. 
If $\FORCE{e_x}$ holds, then we eagerly 
force evaluation of $e_x$ (via the first rule),
otherwise we $\beta$-reduce by substituting
$e_x$ inside the body $e$ (via the second rule).
Note that if $e_x$ is already a value, then 
trivially, only the second rule can be applied.
The same idea generalizes to the \emph{let} and \emph{fix} cases.
Thus, we get lazy (call-by-name) and eager
(call-by-value) semantics by
instantiating $\R$ appropriately
$$\begin{array}{rcll}
\FORCE{e} & \Leftrightarrow & \mathit{false}        & \mbox{\emph{Call-By-Name}} \\
\FORCE{e} & \Leftrightarrow & e\ \mbox{not a value} & \mbox{\emph{Call-By-Value}} \\
\end{array}
$$
We write $\stepv$ (resp. $\stepn$) for the small-step 
relation obtained via the CBV and CBN instantiations of
$\R$.
We write $\evalGen{e}{e'}{}{j}$ (resp. $\evalGen{e}{e'}{}{*}$) 
if $e$ reduces to $e'$ in at most $j$ (resp. finitely many) steps.

\spara{Serious and Trivial Expressions}
Following Reynolds~\cite{Reynolds72}, 
we say that an expression $e$ is 
\emph{trivial} if $\evalns{e}{v}$ 
for some value $v \not = \CRASH$. 
otherwise we call the expression \emph{serious}.
That is, an expression is trivial \emph{iff}
it reduces to a non-$\CRASH$ value under CBN.
Note the set of trivial expressions is closed under
evaluation, \ie if $e$ is trivial and $\eval{e}{e'}$
(for any definition of $\R$) then $e'$ is 
also trivial.

\spara{Optimistic Evaluation}
We define \emph{optimistic evaluation} as 
the small-step relation $\stepo$ obtained
by defining
$$\FORCE{e} \Leftrightarrow e\ \mbox{is trivial} 
			\land e\ \mbox{not a value}$$

In \S~\ref{sec:typing} we will
prove soundness of refinement typing 
with respect to optimistic evaluation. 
To show soundness under lazy evaluation,
we additionally proved~\cite{lazytechrep} the 
following theorem that relates lazy 
and optimistic evaluation.

\includeProof{
\begin{lemma}[Optimistic Reduction]\label{lem:opteval}
If $\evalGen{e}{c}{n}{j}$ then $\evalos{e}{c}$.
\end{lemma}

\proofsketch
We prove the above by induction on $j$, using 
Church-Rosser property,
the definition of trivial and the fact that for 
any closed $e_x$, if $\evalGen{C[e_x]}{v}{n}{k}$ and 
$\evalns{e_x}{v_x}$ then $\evalGen{C[v_x]}{v'}{n}{k}$.
}

\begin{theorem}[Optimistic Equivalence]\label{th:optequiv}
$\evalns{e}{c} \Leftrightarrow \evalos{e}{c}$.
\end{theorem}
\includeProof{
\proofsketch 
The $\Leftarrow$ direction follows from 
Church-Rosser property and 
the fact
that if a term reduces to a value under some 
evaluation strategy, then it cannot diverge under CBN.
The $\Rightarrow$ direction follows from 
Lemma~\ref{lem:opteval}.
}

\section{Type System}\label{sec:typing}

Next, we develop the static semantics for \corelan and 
prove soundness with respect to the OPT, and hence CBN,
evaluation strategies.
We develop the type system in three steps. 
First, we present a general type system where 
termination is tracked abstractly via 
labels
and show how to achieve sound refinement 
typing using a generic \emph{termination oracle}~(\S~\ref{subsec:typing}).
%
%
Next, we describe a concrete instantiation 
%
of the oracle using refinements (\S~\ref{subsec:termination}).
This decoupling allows us to make explicit the
exact requirements for soundness, and also has 
the pragmatic benefit of leaving open the door 
for employing other kinds of termination analyses.


\spara{Basic Types}
Figure~\ref{fig:syntax} summarizes the syntax of \corelan types.
Basic types in \corelan are natural numbers \tint and booleans \tbool. 
We include only two basic types for simplicity; it is straightforward to 
add any type whose values can be sized, and hence ordered using a
well-founded relation, as discussed in~\S~\ref{sec:overview:termination}.

\spara{Labels}
We use two labels to distinguish terminating 
and diverging terms. 
Intuitively, the label \ltop appears in types 
that describe expressions that always terminate, 
while the label \lbot implies that the expressions 
may not terminate.

\spara{Types} 
Types in \corelan include the standard basic refinement type 
$b$ annotated with a label. 
The type \tref{v}{b}{l}{e} describes expressions of type $b$ 
that satisfy the refinement $e$. 
Intuitively, if the label $l$ is \ltop then the expression 
definitely terminates, otherwise it may diverge.
Finally, \corelan types include dependent function types
$\tfun{x}{\tau}{\tau}$.

\spara{Safety}
We formalize safety properties by giving 
various primitive constants the appropriate 
refinement types. For example,
$$\begin{array}{lcl}
\mathtt{3}   & : & \tref{v}{\tint}{\ltop}{v=3} \\
\mathtt{(+)} & : & 
\tfun{x}{\tint^\ltop}{\tfun{y}{\tint^\ltop}{\tref{v}{\tint}{\ltop}{v=x+y}}} \\
\mathtt{(/)} & : & \tfun{x}{\tint^\ltop}
                        {{\tref{v}{\tint}{\ltop}{v \not = 0}} \rightarrow
                              {\tint^\ltop}} \\
\mathtt{error}_\tau & : & \tref{v}{\tint}{\ltop}{\efalse} \rightarrow \tau \\ 
\end{array}$$
We assume that for any constant $c$ with type 
$${\constty{c} \defeq \tfun{x}{\tref{v}{b}{l}{e}}{\tau}}$$ 
for any value $v$, 
if the formula $e$ is valid then the value $\dbrkts{c}(v)$ 
is not equal to $\CRASH$ and can be typed as $\tau$.
Thus, we define \emph{safety} by requiring that
a term does not evaluate to $\CRASH$.

\spara{Serious and Trivial Types} 
A type $\tau$ is \serious if 
$\tau \defeq \tref{v}{b}{\lbot}{e}$,
otherwise, it is \trivial.

\spara{Notation} 
We write $b^l$ to abbreviate the \emph{unrefined}
type \hbox{\tref{v}{b}{l}{\etrue}}.
We ensure that \serious types
(which are, informally speaking, assigned to 
potentially diverging terms) are unrefined.
We write $b$ for $b^l$, when label $l$
can be either \ltop or \lbot.


\subsection{Type-checking}\label{subsec:typing}
\begin{figure}[t!]
\judgementHead{Well-Formedness}{\iswellformed{\Gamma}{\tau}}
$$
\inference{
	\Gamma' = \strict(\Gamma, \tbind{v}{b^\ltop}) &&
	\hastype{\Gamma'}{e}{\tbool}
}{
	\iswellformed{\Gamma}{\tref{v}{b}{\ltop}{e}}
}[\rwbasetop]
$$

$$\begin{array}{cc}
\inference{
}{
	\iswellformed{\Gamma}{\tref{v}{b}{\lbot}{\etrue}}
}[\rwbasebot]
&
\inference{
	\iswellformed{\Gamma}{\tau} \ \
	\iswellformed{\Gamma, \tbind{x}{\tau}}{\tau'}
}{
	\iswellformed{\Gamma}{\tfun{x}{\tau}{\tau'}}
}[\rwfun]
\end{array}$$

\judgementHead{Subtyping}{\issubtype{\Gamma}{\tau_1}{\tau_2}}

$$
\inference{
	\smtvalid{\embed{\Gamma} \Rightarrow \embed{e_1} \Rightarrow \embed{e_2}}
}{
	\issubtype{\Gamma}{\tref{v}{b}{\ltop}{e_1}}{\tref{v}{b}{\ltop}{e_2}}
}[\rsbasetop]
$$

$$
\inference{
}{
	\issubtype{\Gamma}{\tref{v}{b}{l}{e}}{\tref{v}{b}{\lbot}{\etrue}}
}[\rsbasebot]
$$

$$
\inference{
	\issubtype{\Gamma}{\tau_2}{\tau_1} &&
	\issubtype{\Gamma, \tbind{x}{\tau_2}}{\tau_1'}{\tau_2'}
}{
	\issubtype{\Gamma}{\tfun{x}{\tau_1}{\tau_1'}}
	                  {\tfun{x}{\tau_2}{\tau_2'}}
}[\rsfun]
$$

\judgementHead{Typing}{\hastype{\Gamma}{e}{\tau}}
$$
\inference{
	\hastype{\Gamma}{e}{\tau_1} && 
	\issubtype{\Gamma}{\tau_1}{\tau_2} &&
	\iswellformed{\Gamma}{\tau_2}
}{
	\hastype{\Gamma}{e}{\tau_2}
}[\rtsub]
$$

$$\begin{array}{cc}
\inference{}
          {\hastype{\Gamma}{c}{\constty{c}}}[\rtconst]
&
\inference{
	\tbind{x}{\tref{v}{b}{\ltop}{e}} \in \Gamma
}{
	\hastype{\Gamma}{x}{\tref{v}{b}{\ltop}{v = x}}
}[\rtvara]
\end{array}$$

$$
\inference{
	\tbind{x}{\tau} \in \Gamma  && 
	\tau \neq \tref{v}{b}{\ltop}{e}
}{
	\hastype{\Gamma}{x}{\tau}
}[\rtvarb]
$$


$$
\inference{
	\hastype{\Gamma, \tbind{x}{\tau_x}}{e}{\tau} &&
	\iswellformed{\Gamma}{\tfun{x}{\tau_x}{\tau}}
}{
	\hastype{\Gamma}{(\efun{x}{\tau_x}{e})}{\tfun{x}{\tau_x}{\tau}}
}[\rtfun]
$$


$$
\inference{
	\hastype{\Gamma}{e_1}{(\tfun{x}{\tau_x}{\tau})} &
	\hastype{\Gamma}{e_2}{\tau_x} 
}{
	\hastype{\Gamma}{\eapp{e_1}{e_2}}{\SUBST{\tau}{x}{e_2}}
}[\rtapp]
$$

$$
\inference{
	\hastype{\Gamma}{e_x}{\tau_x} &&
	\hastype{\Gamma, \tbind{x}{\tau_x}}{e}{\tau} &&
	\iswellformed{\Gamma}{\tau}
}{
	\hastype{\Gamma}{\elet{x}{e_x}{e}}{\tau}
}[\rtlet]
$$
$$
\inference{
	\hastype{\Gamma, \tbind{x}{\tau_x}, \tbind{f}{\tfun{x}{\tau_x}{\tau}}}
			{e}{\tau} &&
	\iswellformed{\Gamma}{\tfun{x}{\tau_x}{\tau}}
}{
	\hastype{\Gamma}{\erec{f}{x}{e}}{\tfun{x}{\tau_x}{\tau}}
}[\rtrec]
$$
\label{fig:typing}
\caption{Type-checking for \corelan}
\end{figure}

Next, we present the static semantics of \corelan by
describing the type-checking judgments and rules. 
A type environment $\Gamma$ is a sequence of type
bindings $\tbind{x}{\tau}$. 
We use environments to define three kinds of 
rules: Well-formedness, Subtyping and Typing,
which are mostly
standard~\cite{Knowles10,GordonTOPLAS2011}.
The changes are that \serious types are 
unrefined, and binders with \serious types cannot 
appear in refinements.

\spara{Well-formedness}
A judgment \iswellformed{\Gamma}{\tau} states that 
the refinements in $\tau$ are boolean values in the 
environment $\Gamma$ restricted only to the binders whose
types are \emph{\trivial}.
The key rule is \rwbasetop, which checks that the
refinement $e$ of a \trivial basic type is a 
boolean value under the environment $\Gamma'$ 
which contains all of the \trivial bindings in
$\Gamma$ extended with the binding $v:b^\ltop$.
To get the \trivial bindings we use the function 
$\strict$ defined as:
$$\begin{array}{lcll}
\strict(\emptyset) & \defeq & \emptyset \\
\strict(\tbind{x}{\tau}, \Gamma) & \defeq & \tbind{x}{\tau}, \strict(\Gamma) & \text{if}\ \tau\ \text{is \trivial}\\
\strict(\tbind{x}{\tau}, \Gamma) & \defeq & \strict(\Gamma)             & \text{otherwise} \\
\end{array}$$
The rule \rwbasebot ensures \serious types are unrefined.

\spara{Subtyping} 
A judgment \issubtype{\Gamma}{\tau_1}{\tau_2} states 
that the type $\tau_1$ is a subtype of the type 
$\tau_2$ under environment $\Gamma$.
That is, informally speaking, when the free variables 
of $\tau_1$ and $\tau_2$ are bound to values 
described by $\Gamma$, the set of values described
by $\tau_1$ is contained in the set of values 
described by $\tau_2$. 
The interesting rules are the two rules that check 
subtyping of basic types.
Rule \rsbasebot checks subtyping on \serious basic 
types, and requires that the supertype is unrefined.
Rule \rsbasetop checks subtyping on \trivial basic types. 
As usual, subtyping reduces to implication checking:
the \emph{embedding} of the environment $\Gamma$
strengthened with the interpretation of $e_1$ in the
logic should imply the interpretation of $e_2$ in 
the refinement logic.
%
%
The crucial difference is in the definition of embedding.
Here, we keep only the \trivial binders:
$$
\embed{\Gamma} \defeq 
	\bigwedge \{ \embed{\SUBST{e}{v}{x}} \mid
    \tbind{x}{\tref{v}{b}{l}{e}}
    \in \strict(\Gamma)\}
$$
%

\spara{Typing}
A judgment \hastype{\Gamma}{e}{\tau} states that
the expression $e$ has the type $\tau$ 
under environment $\Gamma$.
That is, when the free variables in $e$ are bound 
to values described by $\Gamma$, the expression 
$e$ will evaluate to a value described by $\tau$.
Most of the rules are standard, we discuss only
the interesting ones.

A variable expression $x$ has a type if a binding
$\tbind{x}{\tau}$ exists in the environment.
The rule that is used for typing depends on the 
structure of $\tau$.
If $\tau$ is basic and \trivial, 
the rule \rtvara is used which,
as usual, refines the  basic type 
with the \emph{singleton} refinement 
$v = x$~\cite{Ou2004}.
Otherwise, the rule \rtvarb is used 
to type $x$ with $\tau$.
If $\tau$ is basic and \serious it
should be unrefined, so its type
is \emph{not} strengthened with the 
singleton refinement.
	

The dependent application rule \rtapp is standard:
the type of the expression \eapp{e_1}{e_2}
is $\tau$ where $x$ is 
replaced with the expression $e_2$.
Note that if $\tau_x$ is \serious then $e_2$ 
may diverge and so should not appear in any type. 
In this case, well-formedness ensures that
$x$ will not appear inside $\tau$, and so 
$\tau\sub{x}{e_2} = \tau$ (which does not contain $e_2$.)
Our system allows \trivial arguments to be passed 
to a function that expects a \serious input, but 
not the other way around.
%

\spara{Soundness With a Termination Oracle}
We prove soundness via preservation
and progress theorems with respect to OPT evaluation,
assuming the existence of a \emph{termination oracle}
that assigns labels to types so that \serious 
expressions get \serious types.

\begin{hypothesis}[Termination Oracle]\label{hyp:oracle}
If \hastype{\emptyset}{e}{\tau} and $\tau$ is
a \trivial type, then $e$ is a \trivial expression.
\end{hypothesis}

The proof can be found in~\cite{lazytechrep}, 
and relies on two facts:
(1)~We define constants so that 
the application of a constant
is defined for every value (as required for progress)
but preserves typing for values that satisfy their
input refinements (as required for preservation).
(2)~Variables with \serious types
do not appear in any refinements. 
In particular, they do not appear in any type or environment;
which makes it safe to trivially substitute them with any 
expression whilst preserving typing.
With this, we can prove that 
application of a \serious expressions preserves typing.

We combine preservation and progress to obtain the following result.
The \emph{crash-freedom} guarantee states that if a term $e$ is 
well-typed, it will not crash under \emph{call-by-name} semantics. 
From preservation and progress (and the fact that the 
constant \CRASH has no type), we can show that a well-typed 
term will not crash under \emph{optimistic} evaluation 
($\stepo$). Hence, we can use Theorem~\ref{th:optequiv} 
to conclude that the well-typed term cannot \CRASH under 
\emph{lazy} evaluation ($\stepn$).
Similarly, type-preservation is translated from $\stepo$ to $\stepn$ 
by  observing that a term reduces to a constant under $\stepo$ iff it
reduces to the same constant under $\stepn$.


\begin{theorem}[Safety]\label{th:oracle}
Assuming the Termination Hypothesis,
\begin{itemize}
\item\textbf{Crash-Freedom:} If \hastypet{\emptyset}{e}{\tau} 
        then $e \not \stepn^{*} \CRASH$.
\item\textbf{Type-Preservation:} If \hastypet{\emptyset}{e}{\tau} and
       $\evalns{e}{c}$ then $\hastypet{\emptyset}{c}{\tau}$.
\end{itemize}
\end{theorem}

\subsection{Termination Analysis}\label{subsec:termination}

\begin{figure}
$$
\inference{
	\tau \text{ is \serious} &&
	\iswellformed{\Gamma}{\tfun{x}{\tau_x}{\tau}}\\
	\hastype{\Gamma, x:\tau_x, f : \tfun{y}{\tau_x}{\tau}}{e}{\tau}
}{
	\hastype{\Gamma}{\erec{f}{x}{e}}{\tfun{x}{\tau_x}{\tau}}
}[\rtrecs]
$$
$$
\inference{
	\tau \text{ is \trivial} &&
	\iswellformed{\Gamma}{\tfun{x}{\tau_x}{\tau}} \\
	\tau_x = \tref{v}{\tint}{\ltop}{e_x} &&
	\tau_y = \tref{v}{\tint}{\ltop}{e_x \land v < x} \\
	\hastype{\Gamma, x:\tau_x, f : \tfun{y}{\tau_y}{\tau\sub{x}{y}}}{e}{\tau}
}{
	\hastype{\Gamma}{\erec{f}{x}{e}}{\tfun{x}{\tau_x}{\tau}}
}[\rtrect]
$$
\caption{Typing rules for Termination}
\label{fig:rec}
\end{figure}

Finally, we present an instantiation of the termination oracle 
that essentially generalizes the notion of sized 
types~\cite{HughesParetoSabry96,Sereni05,BartheTermination}
to the refinement setting, and hence complete the description 
of a sound refinement type system for \corelan.
As described in \S~\ref{sec:overview}, the key idea is to
change the rule for typing recursive functions, so that 
the body of the function is checked under a termination-weakened
environment where the function can only be called
on \emph{smaller} inputs. 

\spara{Termination-Weakened Environments}
We formalize this idea by spliting the rule \rtrec that
types fixpoints into the two rules shown in Figure~\ref{fig:rec}. 
The rule \rtrecs can only be applied when the output type 
$\tau$ is \serious, meaning that a recursive function typed
with this rule can (when invoked) diverge.
In contrast, rule \rtrect is used to type a recursive
function that always terminates, \ie whose output type is 
trivial.
The rule requires the argument $x$ of such a function
to be a trivial (terminating) natural number.
Furthermore, when typing the body $e$, the type
of the recursive function $f : \tfun{y}{\tau_y}{\tau}$ 
is \emph{weakened} to enforce that
the function is invoked with an argument $y$ that is 
strictly smaller than $x$ at each recursive call-site.
For clarity of exposition, we require that the 
decreasing metric be the value of the first 
parameter. 
It is straightforward to generalize the requirement 
to any well-founded metric on the arguments.

We prove that our modified type system satisfies 
the termination oracle hypothesis.
That is, in \corelan trivial types are only
ascribed to trivial expressions.
This discharges the Termination Hypothesis 
needed by Theorem~\ref{th:oracle}
and proves safety of \corelan.

\begin{theorem}[Termination]\label{th:termination}
If \hastypet{\emptyset}{e}{\tau} and $\tau$ is \trivial then
$e$ is trivial.
\end{theorem}


The full proof is in~\cite{lazytechrep}, here we summarize the key parts.

\spara{Well-formed Terms}
We call a term $e$ \emph{well-formed} with respect to a type $\tau$,
written $\wfmodels{\tau}{e}$, if:
(1)~if $\tau\equiv\tref{v}{b}{\ltop}{e'}$ then $\evalns{e}{v}$, for some value $v$,
and,
(2)~if $\tau\equiv\tfun{x}{\tau'_x}{\tau'}$ 
then $\evalns{e}{v}$, for some value $v$,
and for all $e_x$ such that $\hastypet{\emptyset}{e_x}{\tau'_x}$
and $\wfmodels{\tau'_x}{e_x}$, 
we have $\wfmodels{\tau' \sub{x}{e_x}}{\eapp{e}{e_x}}$.

\spara{Well-formed Substitutions}
A \emph{substitution} $\theta$ is either empty ($\emptyset$) or of the form
$\theta;\sub{x}{e}$. A substitution $\theta$ is \emph{well-formed} with respect to
an environment $\Gamma$, written $\wfmodels{\Gamma}{\theta}$,
if either both are empty, or the environment and
the substitution are respectively $\Gamma;\tbind{x}{\tau}$ and
$\SUBST{\theta}{x}{e}$, and
(1)~$\wfmodels{\Gamma}{\theta}$,
(2)~$\hastypet{\emptyset}{\theta e}{\theta \tau}$, and
(3)~$\wfmodels{\theta\tau}{\theta e}$.
Now, we can connect refinement types and termination.

\includeProof{
Informally, a well-formed substitution is an assignment
of terms to binders that denotes a valid \emph{closing} of 
an expression with respect to an environment. 
}

\includeProof{
\begin{lemma}[Substitution]\label{lemma:vssub}
If $\wfmodels{\Gamma}{\theta}$ and $\hastypet{\Gamma}{e}{\tau}$, 
then $\hastypet{\emptyset}{\theta e}{\theta \tau}$.
\end{lemma}

\noindent
The lemma follows from (strengthened versions of) lemmas~\ref{lemma:vsub}, \ref{lemma:ssub}.
}

\begin{lemma}[Termination]\label{lemma:termination}
If $\hastype{\Gamma}{e}{\tau}$ and $\wfmodels{\Gamma}{\theta}$,
then $\wfmodels{\theta \tau}{\theta e}$.
\end{lemma}

The Termination Theorem~\ref{th:termination} is an 
immediate corollary of Lemma~\ref{lemma:termination} where
$\Gamma$ is the empty environment.
\includeProof{
\noindent
We prove this by induction on the typing derivation 
$\hastypet{\Gamma}{e}{\tau}$.
The interesting case is when \rtrect is used, \ie for a \trivial $\tau$ 
$$\hastype{\Gamma}{\erec{f}{x}{e'}}{\tfun{x}{\tau_x}{\tau}}$$
where $e \defeq \erec{f}{x}{e'}$.
In this case, we need to prove the goal that for any well-formed 
substitution $\theta$, \ie $\wfmodels{\Gamma}{\theta}$,
and any expression $e_x$ such that
(a)~$\hastypet{\emptyset}{e_{x}}{\theta \tau_{x}}$ and
(b)~$\wfmodels{\theta \tau_x}{e_{x}}$,
there exists a value $v$ such that $\evalns{{(\theta e)}{e_{x}}}{v}$.
By inverting the typing rule we get 
(c)~$\tau_x \defeq \tref{v}{\tint}{\ltop}{e_x'}$,
(d)~$\tau_y \defeq \tref{v}{\tint}{\ltop}{e_x' \land v < x}$, and
(e)~$\hastypet{\Gamma, \tbind{x}{\tau_x}, \tbind{f}{\tau_f}}{e}{\tau}$
where $\tau_f \defeq \tfun{y}{\tau_y}{\tau\sub{x}{y}}$.
Let $N$ be the set of \emph{values} described by $\theta \tau_x$ 
(\ie that satisfy the refinement).
As $N \subseteq \ttNat$, we can enumerate it as 
$N \defeq n_0,n_1,\ldots,n_i,\ldots$. 
Now, we prove by induction on $i$, that for each $n_i$
there exists a value $v$ such that $\evalns{{(\theta e)}{n_i}}{v}$.
Finally, (by safety and (b)) we conclude 
that $\evalns{e_x}{n_i}$ for some $i$, which concludes the proof.
}
Since we are using refinements to prove termination, 
our termination proof requires soundness and vice versa.
We resolve this circularity by proving
Preservation, Progress, and the Termination Lemma
\includeProof{Lemmas~\ref{lemma:vsub},~\ref{lemma:ssub},~\ref{lemma:vssub} and \ref{lemma:termination}}
by mutual induction, to obtain Theorem~\ref{th:oracle}
\emph{without} the Termination Hypothesis~\cite{lazytechrep}.

\section{Evaluation}\label{sec:evaluation}

We implemented our technique by extending \toolname~\cite{vazou13}. 
Next, we describe the tool, the benchmarks, 
and a quantitative summary of our results.
We then present a qualitative discussion 
of how \toolname was used to verify safety, 
termination, and functional correctness 
properties of a large library, and 
discuss the strengths and limitations 
unearthed by the study.

\spara{Implementation} \toolname takes as input:
(1)~A Haskell \emph{source} file, 
(2)~Refinement type \emph{specifications}, 
    including refined datatype definitions, 
    measures, predicate and type aliases,
    and function signatures, and
(3)~Predicate fragments called \emph{qualifiers} 
    which are used to infer refinement types using 
    the abstract interpretation framework of Liquid
    typing~\cite{LiquidPLDI08}.
The verifier returns as output, \textsc{Safe} or \textsc{Unsafe}, 
depending on whether the code meets the specifications or not, 
and, importantly for debugging the code (or specification!) 
the inferred types for all sub-expressions.

\subsection{Benchmarks and Results}

Our goal was to use \toolname to verify a suite of real-world Haskell 
programs, to evaluate whether our approach is
%
    \emph{efficient}  enough for large programs,
    \emph{expressive} enough to  specify key correctness properties, and
    \emph{precise}    enough to  verify idiomatic Haskell codes.
%
Thus, we used these libraries as benchmarks:
\begin{itemize}
\item \texttt{GHC.List} and \texttt{Data.List}, which together implement many standard
      list operations; we verify various
      size related properties,
\item \texttt{Data.Set.Splay}, which implements a splay-tree
      based functional set data type; we verify that all interface 
      functions terminate and return well ordered trees,
\item \texttt{Data.Map.Base}, which implements a functional 
      map data type; we verify that all interface functions 
      terminate and return binary-search ordered trees~\cite{vazou13}, 
\item \libvectoralgos, which includes a suite of 
      ``imperative" (\ie monadic) array-based sorting algorithms; 
      we verify the correctness of vector accessing, indexing, and slicing.
\item \bytestring, a library for manipulating byte arrays, we
      verify termination, low-level memory safety, and high-level
      functional correctness properties\includeProof{(\ref{sec:bytestring})},
\item \libtext, a library for high-performance unicode text 
      processing; we verify various pointer safety and 
      functional correctness properties (\ref{sec:text}),
      during which we find a subtle bug.
\end{itemize}
We chose these benchmarks as they represent a wide spectrum of idiomatic
Haskell codes: the first three are widely used libraries based on recursive 
data structures, the fourth and fifth perform subtle, low-level arithmetic 
manipulation of array indices and pointers, and the last is a rich, high-level
library with sophisticated application-specific invariants. 
These last three libraries are especially representative as they pervasively 
intermingle high level abstractions like higher-order loops, folds, and fusion, 
with low-level pointer manipulations in order to deliver high-performance. 
They are an appealing target for \toolname, as refinement types are an ideal 
way to statically enforce critical invariants that are outside the scope of
run-time checking as even Haskell's highly expressive type system.

\begin{table*}
\begin{scriptsize}
\centering
\begin{tabular}{|l|rrrr|rrrr|r|}
\hline
\bfModule & \bfLOC & \bfSpecs 
& \bfAnnot & \bfQualif & \bfLetRec & \bfSerious
& \bfHint & \bfWitness & \textbf{Time (s)} \\
\hline \hline
\texttt{GHC.List}                            & 310   & 29 / 34    & 8 / 8     & 3 / 3     & 37  & 5  & 1  & 0  & 23 \\
\texttt{Data.List}                           & 504   & 10 / 18    & 7 / 7     & 0 / 0     & 50  & 2  & 5  & 3  & 38 \\
\hline
\texttt{Data.Set.Splay}                      & 149   & 27 / 37    & 5 / 5     & 0 / 0     & 17  & 0  & 4  & 3  & 25 \\
\hline
\texttt{Data.Map.Base}                       & 1396  & 123 / 173  & 12 / 12   & 0 / 0     & 88  & 0  & 9  & 2  & 219 \\

\hline
\texttt{Data.ByteString}                     & 1035  & 94 / 117   & 10 / 10   & 5 / 10    & 49  & 2  & 7  & 21 & 249 \\
\texttt{Data.ByteString.Char8}               & 503   & 20 / 26    & 5 / 5     & 0 / 0     & 9   & 0  & 5  & 2  & 19 \\
\texttt{Data.ByteString.Fusion}              & 447   & 27 / 42    & 0 / 0     & 8 / 8     & 23  & 0  & 0  & 6  & 55 \\
\texttt{Data.ByteString.Internal}            & 272   & 44 / 101   & 1 / 1     & 10 / 10   & 5   & 0  & 0  & 0  & 12 \\
\texttt{Data.ByteString.Lazy}                & 673   & 69 / 89    & 24 / 24   & 24 / 96   & 59  & 4  & 20 & 1  & 338 \\
\texttt{Data.ByteString.Lazy.Char8}          & 406   & 13 / 19    & 4 / 4     & 0 / 0     & 7   & 0  & 4  & 0  & 20 \\
\texttt{Data.ByteString.Lazy.Internal}       & 55    & 18 / 43    & 2 / 2     & 0 / 0     & 8   & 0  & 0  & 0  & 2 \\
\texttt{Data.ByteString.Unsafe}              & 110   & 13 / 19    & 0 / 0     & 0 / 0     & 0   & 0  & 0  & 0  & 3 \\

\hline
\texttt{Data.Text}                           & 801   & 73 / 153   & 6 / 6     & 1 / 5     & 35  & 0  & 3  & 11 & 221 \\
\texttt{Data.Text.Array}                     & 167   & 35 / 87    & 2 / 2     & 7 / 9     & 1   & 0  & 0  & 1  & 7 \\
\texttt{Data.Text.Encoding}                  & 189   & 8 / 25     & 1 / 1     & 8 / 10    & 4   & 1  & 0  & 3  & 197 \\
\texttt{Data.Text.Foreign}                   & 84    & 7 / 11     & 0 / 0     & 2 / 2     & 7   & 0  & 0  & 2  & 6 \\
\texttt{Data.Text.Fusion}                    & 177   & 2 / 2      & 3 / 3     & 9 / 22    & 5   & 3  & 0  & 0  & 106 \\
\texttt{Data.Text.Fusion.Size}               & 112   & 10 / 22    & 1 / 1     & 2 / 2     & 7   & 0  & 0  & 0  & 6 \\
\texttt{Data.Text.Internal}                  & 55    & 22 / 49    & 10 / 12   & 9 / 23    & 0   & 0  & 0  & 0  & 5 \\
\texttt{Data.Text.Lazy}                      & 801   & 74 / 150   & 12 / 12   & 1 / 5     & 50  & 0  & 12 & 1  & 215 \\
\texttt{Data.Text.Lazy.Builder}              & 147   & 6 / 31     & 1 / 1     & 2 / 4     & 4   & 0  & 1  & 1  & 28 \\
\texttt{Data.Text.Lazy.Encoding}             & 127   & 2 / 2      & 0 / 0     & 0 / 0     & 4   & 0  & 0  & 0  & 11 \\
\texttt{Data.Text.Lazy.Fusion}               & 84    & 4 / 11     & 1 / 1     & 0 / 0     & 3   & 1  & 0  & 0  & 9 \\
\texttt{Data.Text.Lazy.Internal}             & 66    & 19 / 38    & 5 / 5     & 2 / 4     & 5   & 0  & 0  & 0  & 3 \\
\texttt{Data.Text.Lazy.Search}               & 104   & 13 / 68    & 5 / 5     & 0 / 0     & 6   & 0  & 5  & 0  & 161 \\
\texttt{Data.Text.Private}                   & 25    & 2 / 4      & 0 / 0     & 0 / 0     & 1   & 0  & 0  & 1  & 2 \\
\texttt{Data.Text.Search}                    & 61    & 5 / 15     & 0 / 0     & 2 / 2     & 4   & 0  & 0  & 3  & 20 \\
\texttt{Data.Text.Unsafe}                    & 74    & 10 / 33    & 5 / 5     & 2 / 7     & 0   & 0  & 0  & 0  & 6 \\
\texttt{Data.Text.UnsafeChar}                & 51    & 8 / 11     & 0 / 0     & 2 / 2     & 0   & 0  & 0  & 0  & 4 \\

\hline
\texttt{Data.Vector.Algorithms.AmericanFlag} & 270   & 7 / 17     & 2 / 2     & 3 / 3     & 7   & 0  & 2  & 7  & 44 \\
\texttt{Data.Vector.Algorithms.Combinators}  & 26    & 0 / 0      & 0 / 0     & 0 / 0     & 0   & 0  & 0  & 0  & 0 \\
\texttt{Data.Vector.Algorithms.Common}       & 33    & 25 / 82    & 0 / 0     & 9 / 9     & 1   & 0  & 0  & 0  & 3 \\
\texttt{Data.Vector.Algorithms.Heap}         & 170   & 11 / 31    & 1 / 1     & 0 / 0     & 4   & 0  & 1  & 2  & 53 \\
\texttt{Data.Vector.Algorithms.Insertion}    & 51    & 4 / 14     & 0 / 0     & 0 / 0     & 2   & 0  & 0  & 1  & 4 \\
\texttt{Data.Vector.Algorithms.Intro}        & 142   & 13 / 35    & 3 / 3     & 0 / 0     & 6   & 0  & 3  & 2  & 26 \\
\texttt{Data.Vector.Algorithms.Merge}        & 71    & 0 / 0      & 3 / 3     & 1 / 1     & 4   & 0  & 3  & 4  & 25 \\
\texttt{Data.Vector.Algorithms.Optimal}      & 183   & 3 / 12     & 0 / 0     & 0 / 0     & 0   & 0  & 0  & 0  & 36 \\
\texttt{Data.Vector.Algorithms.Radix}        & 195   & 4 / 16     & 0 / 0     & 0 / 0     & 5   & 0  & 0  & 3  & 8 \\
\texttt{Data.Vector.Algorithms.Search}       & 78    & 3 / 15     & 0 / 0     & 0 / 0     & 3   & 0  & 0  & 3  & 10 \\

\hline
\textbf{Total}                               & 10204 & 857 / 1652 & 139 / 141 & 112 / 237 & 520 & 18 & 85 & 83 & 2240 \\
\hline
\end{tabular}
\caption{\small A quantitative evaluation of our experiments. \bfLOC is the 
number of non-comment lines of source code as reported by \texttt{sloccount}. 
  \bfSpecs  is the number (/ line-count) of type specifications and aliases,
                  data declarations, and measures provided.
  \bfAnnot  is the number (/ line-count) of other annotations provided,
                  these include invariants and hints for
                  the termination checker.
  \bfQualif is the number (/ line-count) of provided qualifiers.
  \bfLetRec is the number of recursive functions in the module.
  \bfSerious is the number of functions marked as potentially non-terminating.
  \bfHint   is the number of termination hints given to \toolname,
                  which specify \emph{which} parameter decreases (by
                  default: the first parameter that has a size metric).
  \bfWitness   is the number of functions that required the addition 
                  of a ghost termination witness parameter. 
  \textbf{Time}   is the time, in seconds, required to run \toolname on the module.}
\label{table:results}
\end{scriptsize}
\end{table*}

\spara{Results}
Table~\ref{table:results} summarizes our experiments, which covered 39 modules
totaling 10204 non-comment lines of source code and 1652 lines of specifications.
The results are on a machine with an Intel Xeon X5660 and 32GB of RAM~(no benchmark required more than 1GB.)
The upshot is that \toolname is very effective on real-world code bases.
The total overhead due to hints, \ie the sum of \bfAnnot, \bfQualif, and \bfWitness (\bfHint is
included in \bfAnnot), is 4.5\% of \bfLOC.
The specifications themselves are machine checkable versions of the comments 
placed around functions describing safe usage and behavior.
Our default metric, namely the first parameter with an associated size measure,
suffices to prove 67\% of (recursive) functions terminating.
30\% require a hint (\ie the position of the decreasing argument) or a
witness (3\% required both), and the remaining 3\% were marked as potentially diverging.
Of the 18 functions marked as potentially diverging, we suspect 6
actually terminate but were unable to prove so.
While there is much room for improving the running times, the tool is fast enough 
to be used interactively, verify a handful of API functions and associated helpers 
in isolation.

\includeBytestring{
\subsection{Bytestring}\label{sec:bytestring}
\input{bytestring}
}

\subsection{Case Study: Text}\label{sec:text}
Next, to give a qualitative sense of the kinds of properties analyzed 
during the course of our evaluation, we present a brief overview of
the verification of \libtext, which is the standard library used for
serious unicode text processing in Haskell.

\libtext uses byte arrays and stream fusion to guarantee 
performance while providing a high-level API.
In our evaluation of \toolname on \libtext~\cite{text},
we focused on two types of properties: 
(1) the safety of array index and write operations, and 
(2) the functional correctness of the top-level API.
These are both made more interesting by the fact that 
\libtext internally encodes characters using UTF-16, 
in which characters are stored in either two or four bytes.
\libtext is a vast library spanning 39 modules and 5700 lines of
code, however we focus on the 17 modules that are relevant
to the above properties.
While we have verified exact functional correctness size properties
for the top-level API, we focus here on the low-level functions 
and interaction with unicode.

\spara{Arrays and Texts}
A @Text@ consists of an (immutable) @Array@ of 16-bit words,
an offset into the @Array@, and a length describing the
number of @Word16@s in the @Text@.  
The @Array@ is created and filled using a
\emph{mutable} @MArray@. 
All write operations in \libtext are performed on @MArray@s 
in the @ST@ monad, but they are \emph{frozen} into @Array@s
before being used by the @Text@ constructor.
We write a measure denoting the size of an @MArray@ and use
it to type the write and freeze operations.
\begin{code}
measure malen       :: MArray s -> Int
predicate EqLen A MA = alen A = malen MA
predicate Ok I A     = 0 <= I < malen A
type VO A            = {v:Int| Ok v A} 

unsafeWrite  :: m:MArray s
             -> VO m -> Word16 -> ST s ()
unsafeFreeze :: m:MArray s
             -> ST s {v:Array | EqLen v m}
\end{code}

\spara{Reasoning about Unicode}
The function @writeChar@ (abbreviating \texttt{UnsafeChar.unsafeWrite})
writes a @Char@ into an @MArray@.
\libtext uses UTF-16 to represent characters internally,
meaning that every @Char@ will be encoded using two or 
four bytes (one or two @Word16@s).
\begin{code}
writeChar marr i c
    | n < 0x10000 = do
        unsafeWrite marr i (fromIntegral n)
        return 1
    | otherwise = do
        unsafeWrite marr i lo
        unsafeWrite marr (i+1) hi
        return 2
    where n = ord c
          m = n - 0x10000
          lo = fromIntegral
             $ (m `shiftR` 10) + 0xD800
          hi = fromIntegral
             $ (m .&. 0x3FF) + 0xDC00
\end{code}
The UTF-16 encoding complicates the specification of the function
as we cannot simply require @i@ to be less than the length of 
@marr@; if @i@ were @malen marr - 1@ and @c@ required two 
@Word16@s, we would perform an out-of-bounds write. 
We account for this subtlety with a predicate that states 
there is enough @Room@ to encode @c@.
%
\begin{code}
predicate OkN I A N  = Ok (I+N-1) A
predicate Room I A C = if ord C < 0x10000
                       then OkN I A 1
                       else OkN I A 2

type OkSiz I A = {v:Nat  | OkN  I A v}
type OkChr I A = {v:Char | Room I A v}
\end{code}
@Room i marr c@ says 
``if @c@ is encoded using one @Word16@, 
  then @i@ must be less than @malen marr@,
  otherwise @i@ must be less than @malen marr - 1@.''
@OkSiz I A@ is an alias for a valid number of @Word16@s 
remaining after the index @I@ of array @A@. 
@OkChr@ specifies the @Char@s for which there is room (to write)
at index @I@ in array @A@.
The specification for @writeChar@ states that given an array @marr@, 
an index @i@, and a valid @Char@ for which there is room at index @i@,
the output is a monadic action returning the number of @Word16@ occupied
by the @char@.
\begin{code}
writeChar :: marr:MArray s
          -> i:Nat
          -> OkChr i marr
          -> ST s (OkSiz i marr)
\end{code}
\spara{Bug}
Thus, clients of @writeChar@ should only call it with suitable indices
and characters.
Using \toolname we found an error in one client, @mapAccumL@, 
which combines a map and a fold over a @Stream@, and stores 
the result of the map in a @Text@. Consider the inner loop of @mapAccumL@.
%
\begin{code}
outer arr top = loop
 where
  loop !z !s !i =
    case next0 s of
      Done          -> return (arr, (z,i))
      Skip s'       -> loop z s' i
      Yield x s'
        | j >= top  -> do
          let top' = (top + 1) `shiftL` 1
          arr' <- new top'
          copyM arr' 0 arr 0 top
          outer arr' top' z s i
        | otherwise -> do
          let (z',c) = f z x
          d <- writeChar arr i c
          loop z' s' (i+d)
        where j | ord x < 0x10000 = i
                | otherwise       = i + 1
\end{code}
Let's focus on the @Yield x s'@ case.
We first compute the maximum index @j@ to 
which we will write and determine the safety of a write. 
If it is safe to write to @j@ we call the provided 
function @f@ on the accumulator @z@ and the character 
@x@, and write the \emph{resulting} character @c@ into the array. 
However, we know nothing about @c@, in particular, 
whether @c@ will be stored as one or two @Word16@s! 
Thus, \toolname flags the call to @writeChar@ as \emph{unsafe}.
The error can be fixed by lifting @f z x@ into the @where@ clause and defining the
write index @j@ by comparing @ord c@ (not @ord x@). \toolname (and the authors)
readily accepted our fix.

\subsection{Code Changes}\label{sec:changes}

Our case studies also highlighted some limitations
of \toolname that we will address in future work. 
In most cases, we could alter the code slightly to 
facilitate verification. 
We briefly summarize the important categories here; 
refer to \cite{lazytechrep} for details.

\spara{Ghost parameters} are sometimes needed in 
order to materialize values that are not needed 
for the computation, but are necessary to prove 
the specification; proving termination may
require a decreasing value that is a function
of several parameters.
In future work it will be interesting to explore
the use of advanced techniques for synthesizing 
ranking witnesses \cite{Terminator} to eliminate
such parameters.

\includeBytestring{
\spara{Higher-order functions} 
must sometimes 
be \emph{specialized} because the original type
is not precise enough. 
For example, the \texttt{concat} function that
concatenates a list of input @ByteString@s 
pre-allocates the output region by computing the 
total size of the input.
\begin{code}
len = sum . map length $ xs
\end{code}
Unfortunately, the type for @map@ is not sufficiently
precise to conclude that the value @len@ equals 
@bLens xs@, se we must manually specialize
the above into a single recursive traversal that 
computes the lengths.
Rather than complicating the type system with a very
general higher-order type for @map@ we suspect the 
best way forward will be to allow the user to specify
inlining in a clean fashion.
}

\spara{Lazy binders} sometimes get in the way of verification. 
A common pattern in Haskell code is to define \emph{all} 
local variables in a single @where@ clause and use them 
only in a subset of all branches. 
\toolname flags a few such definitions as \emph{unsafe},
not realizing that the values will only be demanded in a
specific branch. 
Currently, we manually transform the code by pushing 
binders inwards to the usage site. 
This transformation could be easily automated.

\spara{Assumes} 
which can be thought of as ``hybrid" run-time checks,
had to be placed in a couple of cases where the verifier 
loses information. 
One source is the introduction of assumptions about
mathematical operators that are currently conservatively 
modeled in the refinement logic (\eg that multiplication 
is commutative and associative). 
These may be removed by using more advanced non-linear 
arithmetic decision procedures. 


\section{Related Work}\label{sec:related}


Next we situate our work with closely related lines of research.

\spara{Dependent Types} are the basis of many verifiers, 
or more generally, proof assistants.
In this setting arbitrary terms may appear inside types,
so to prevent logical inconsistencies, and enable
the checking of type equivalence, all terms must
terminate.
``Full" dependently typed systems like Coq~\cite{coq-book}, Agda~\cite{norell07}, and Idris~\cite{Brady13}
typically use various \emph{structural} checks where recursion is allowed on
sub-terms of ADTs to ensure that \emph{all} terms terminate.
%
%
One can \emph{fake} ``lightweight" dependent 
types in Haskell \cite{ChakravartyKJ05,JonesVWW06,Weirich12}.
In this style, the invariants are expressed in 
a restricted~\cite{Jia10} total 
index language and relationships (\eg $x<y$ and $y<z$) 
are combined (\eg $x<z$) by explicitly constructing
a term denoting the consequent from terms 
denoting the antecedents.
On the plus side this ``constructive" approach
ensures soundness. 
It is impossible to witness inconsistencies, 
as doing so triggers diverging computations.
However, it is unclear how easy
it is to use restricted indices with explicitly
constructed relations to verify the complex 
properties needed for large libraries.



\spara{Refinement Types} 
are a form of dependent types where invariants are
encoded via a combination of types and SMT-decidable
logical refinement predicates~\cite{Rushby98,pfenningxi98}. 
Refinement types offer a highly automated means of
verification and have been applied to check
a variety of program properties, including
functional correctness of data structures~\cite{Dunfield07,LiquidPLDI09}, 
security protocols~\cite{GordonTOPLAS2011,Bhargavan13} 
and compilers~\cite{SwamyPOPL13}.
The language of refinements is restricted
to ensure consistency,
however, program variables (binders), or their
singleton representatives \cite{pfenningxi98}, 
\emph{are} crucially allowed in the refinements. 
As discussed in \S~\ref{sec:overview} this 
is sound under \emph{call-by-value} evaluation, 
but under Haskell's semantics any innocent
binder can be potentially diverging, causing 
unsoundness.
Finally, our implementation is based on 
\toolname~\cite{vazou13}, which was unsound 
as it assumed CBV evaluation.

\spara{Size-based Termination Analyses}
have been used to verify termination of
recursive functions, either using the 
``size-change principle''~\cite{BartheTermination, JonesB04},
or via the type system~\cite{HughesParetoSabry96,Sereni05}
by annotating types with size indices
and verifying that the arguments of recursive calls have smaller indices.
In work closely related to ours,
Xi~\cite{XiTerminationLICS01}
encoded sizes via refinement types to prove totality of programs.
What differentiates the above work from ours
is that we do not aim to prove that all expressions converge; 
on the contrary, under a lazy setting diverging expressions
are welcome.
We use size analysis to track diverging terms
in order to exclude them from the logic.

\spara{Static Checkers} 
like ESCJava~\cite{ESCJava} are a classical way 
of verifying correctness through assertions and 
pre- and post-conditions.
One can view Refinement Types as a type-based 
generalization of this approach. 
Classical contract checkers check ``partial" 
(as opposed to ``total") correctness (\ie safety) 
for \emph{eager}, typically first-order, languages
and need not worry about termination.
We have shown that in the lazy setting, even 
``partial" correctness requires proving ``total" 
correctness!
\cite{XuPOPL09} describes a static contract checker for 
Haskell that uses symbolic execution.
The (checker's) termination requires that recursive 
procedures only be unrolled up to some fixed depth.
While this approach removes inconsistencies, it yields 
weaker, ``bounded" soundness guarantees.
Zeno~\cite{ZENO} is another automatic prover for 
Haskell which proves properties by unrolling 
recursive definitions, rewriting, and goal-splitting,
using sophisticated proof-search techniques 
to ensure convergence.
As it is based on rewriting, ``Zeno might loop forever"
when faced with non-terminating functions,
but will not conclude erroneous facts.
Finally, \cite{halo} describes a novel contract 
checking technique that encodes Haskell programs 
into first-order logic. Intriguingly, the paper 
shows how the encoding, which models the denotational
semantics of the code, is \emph{simplified} by 
lazy evaluation.

Unlike the previous contract checkers for Haskell, 
our type-based approach does not rely on heuristics
for unrolling recursive procedures, and instead uses
SMT and abstract interpretation \cite{LiquidPLDI08}
to infer signatures, which we conjecture makes \toolname
more predictable.
Of course, this requires \toolname be provided logical 
qualifiers (predicate fragments) which form the basis
of the analysis' abstract domain. 
In our experience, however, this is not an onerous 
burden as most qualifiers can be harvested from API 
specifications, and the overall workflow is predictable
enough to enable the verification of large, real-world
code bases.

{
\bibliographystyle{plain}
\bibliography{main}

\begin{thebibliography}{10}

\bibitem{BMMR01}
T.~Ball, R.~Majumdar, T.~Millstein, and S.~K. Rajamani.
\newblock Automatic predicate abstraction of {C} programs.
\newblock In {\em PLDI}, 2001.

\bibitem{BartheTermination}
G.~Barthe, M.~J. Frade, E.~Gim{\'e}nez, L.~Pinto, and T.~Uustalu.
\newblock Type-based termination of recursive definitions.
\newblock {\em Mathematical Structures in Computer Science}, 2004.

\bibitem{GordonTOPLAS2011}
J.~Bengtson, K.~Bhargavan, C.~Fournet, A.~D. Gordon, and S.~Maffeis.
\newblock Refinement types for secure implementations.
\newblock {\em ACM TOPLAS}, 2011.

\bibitem{coq-book}
Y.~Bertot and P.~Cast\'eran.
\newblock {\em Coq'Art: The Calculus of Inductive Constructions}.
\newblock Springer Verlag, 2004.

\bibitem{Bhargavan13}
K.~Bhargavan, C.~Fournet, M.~Kohlweiss, A.~Pironti, and P.-Y Strub.
\newblock Implementing tls with verified cryptographic security.
\newblock In {\em IEEE S \& P}, 2013.

\bibitem{Brady13}
Edwin Brady.
\newblock Idris: general purpose programming with dependent types.
\newblock In {\em PLPV}, 2013.

\bibitem{ChakravartyKJ05}
M.~T. Chakravarty, G.~Keller, and S.~L. Peyton-Jones.
\newblock Associated type synonyms.
\newblock In {\em ICFP}, 2005.

\bibitem{Terminator}
B.~Cook, A.~Podelski, and A.~Rybalchenko.
\newblock Termination proofs for systems code.
\newblock In {\em PLDI}, 2006.

\bibitem{Dunfield07}
J.~Dunfield.
\newblock Refined typechecking with {Stardust}.
\newblock In {\em PLPV}, 2007.

\bibitem{Weirich12}
R.~A. Eisenberg and S.~Weirich.
\newblock Dependently typed programming with singletons.
\newblock In {\em Haskell Symposium}, 2012.

\bibitem{Ennals03}
R.~Ennals and S.~Peyton-Jones.
\newblock Optimistic evaluation: An adaptive evaluation strategy for non-strict
  programs.
\newblock In {\em ICFP}, 2003.

\bibitem{ESCJava}
C.~Flanagan, K.R.M. Leino, M.~Lillibridge, G.~Nelson, J.~B. Saxe, and R.~Stata.
\newblock Extended static checking for {Java}.
\newblock In {\em PLDI}, 2002.

\bibitem{SwamyPOPL13}
C.~Fournet, N.~Swamy, J.~Chen, P-{\'E.} Dagand, P-Y. Strub, and B.~Livshits.
\newblock Fully abstract compilation to javascript.
\newblock In {\em POPL}, 2013.

\bibitem{Greenberg12}
M.~Greenberg, B.~C. Pierce, and S.~Weirich.
\newblock Contracts made manifest.
\newblock {\em JFP}, 2012.

\bibitem{HughesParetoSabry96}
J.~Hughes, L.~Pareto, and A.~Sabry.
\newblock Proving the correctness of reactive systems using sized types.
\newblock In {\em POPL}, 1996.

\bibitem{Jia10}
L.~Jia, J.~Zhao, V.~Sj\"oberg, and S.~Weirich.
\newblock Dependent types and program equivalence.
\newblock In {\em POPL}, 2010.

\bibitem{JonesB04}
Neil~D. Jones and Nina Bohr.
\newblock Termination analysis of the untyped lamba-calculus.
\newblock In {\em RTA}, 2004.

\bibitem{LiquidPLDI09}
M.~Kawaguchi, P.~Rondon, and R.~Jhala.
\newblock Type-based data structure verification.
\newblock In {\em PLDI}, 2009.

\bibitem{Knowles10}
K.W. Knowles and C.~Flanagan.
\newblock Hybrid type checking.
\newblock {\em ACM TOPLAS}, 2010.

\bibitem{Mycroft80}
A.~Mycroft.
\newblock The theory and practice of transforming call-by-need into
  call-by-value.
\newblock In {\em ESOP}, 1980.

\bibitem{norell07}
U.~Norell.
\newblock {\em Towards a practical programming language based on dependent type
  theory}.
\newblock PhD thesis, Chalmers, 2007.

\bibitem{text}
B.~O'Sullivan and T.~Harper.
\newblock text-0.11.2.3: An efficient packed unicode text type.
\newblock \url{http://hackage.haskell.org/package/text-0.11.2.3}.

\bibitem{Ou2004}
X.~Ou, G.~Tan, Y.~Mandelbaum, and D.~Walker.
\newblock Dynamic typing with dependent types.
\newblock In {\em IFIP TCS}, 2004.

\bibitem{JonesVWW06}
S.~L. Peyton-Jones, D.~Vytiniotis, S.~Weirich, and G.~Washburn.
\newblock Simple unification-based type inference for {GADTs}.
\newblock In {\em ICFP}, 2006.

\bibitem{PlotkinTCS75}
Gordon Plotkin.
\newblock Call-by-name, call-by-value and the lambda calculus.
\newblock {\em Theoretical Computer Science}, 1975.

\bibitem{Reynolds72}
John~C. Reynolds.
\newblock Definitional interpreters for higher-order programming languages.
\newblock In {\em 25th ACM National Conference}, 1972.

\bibitem{LiquidPLDI08}
P.~Rondon, M.~Kawaguchi, and R.~Jhala.
\newblock Liquid types.
\newblock In {\em PLDI}, 2008.

\bibitem{Rushby98}
J.~Rushby, S.~Owre, and N.~Shankar.
\newblock Subtypes for specifications: Predicate subtyping in pvs.
\newblock {\em IEEE TSE}, 1998.

\bibitem{Sereni05}
D.~Sereni and N.D. Jones.
\newblock Termination analysis of higher-order functional programs.
\newblock In {\em APLAS}, 2005.

\bibitem{ZENO}
W.~Sonnex, S.~Drossopoulou, and S.~Eisenbach.
\newblock Zeno: An automated prover for properties of recursive data
  structures.
\newblock In {\em TACAS}, 2012.

\bibitem{lazytechrep}
From {Safety} to {Termination} and back: {SMT-Based} verification for {Lazy}
  languages.
\newblock Supplementary Materials.

\bibitem{Turing36}
A.~M. Turing.
\newblock On computable numbers, with an application to the
  eintscheidungsproblem.
\newblock In {\em LMS}, 1936.

\bibitem{TerauchiPOPL13}
H.~Unno, T.~Terauchi, and N.~Kobayashi.
\newblock Automating relatively complete verification of higher-order
  functional programs.
\newblock In {\em POPL}, 2013.

\bibitem{vazou13}
N.~Vazou, P.~Rondon, and R.~Jhala.
\newblock Abstract refinement types.
\newblock In {\em ESOP}, 2013.

\bibitem{halo}
D.~Vytiniotis, S.L. Peyton-Jones, K.~Claessen, and D.~Ros{\'e}n.
\newblock Halo: haskell to logic through denotational semantics.
\newblock In {\em POPL}, 2013.

\bibitem{WadlerICFP03}
P.~Wadler.
\newblock Call-by-value is dual to call-by-name.
\newblock In {\em ICFP}, 2003.

\bibitem{XiTerminationLICS01}
H.~Xi.
\newblock Dependent types for program termination verification.
\newblock In {\em LICS}, 2001.

\bibitem{pfenningxi98}
H.~Xi and F.~Pfenning.
\newblock Eliminating array bound checking through dependent types.
\newblock In {\em PLDI}, 1998.

\bibitem{XuPOPL09}
Dana~N. Xu, Simon~L. Peyton-Jones, and Koen Claessen.
\newblock Static contract checking for haskell.
\newblock In {\em POPL}, 2009.

\end{thebibliography}
}

\includeProof{
	\newpage
	\appendix

	\renewcommand\NV[1]{}

	\input{../proofs/intro}

	\input{../proofs/eval}

	\renewcommand\eval[2]{\ensuremath{{#1} \hookrightarrow _o {#2}}}
	\input{../proofs/soundness}
	\input{../proofs/constants}
	\input{../proofs/anf}
	\input{../proofs/substitutions}
	\input{../proofs/narrowing}
	\input{../proofs/serioussubst}
	\input{../proofs/valuesubst}
	\input{../proofs/preservation}
	\input{../proofs/progress}

	\input{../proofs/termination}

	\section{Case Study: Bytestring}\label{bench:bytestring}
	\input{bytestring}

}

\end{document}